# Ambient Temperature Growth and Superconducting Properties of Ti-V Alloy Thin Films

Shekhar Chandra Pandey[1, 2, a)], Shilpam Sharma[1], Ashish Khandelwal[1] and M. K. Chattopadhyay[1, 2]

[1]*Free Electron Laser Utilization Laboratory, Raja Ramanna Centre for Advanced Technology, Indore, Madhya Pradesh - 452 013, India*
[2]*Homi Bhabha National Institute, Training School Complex, Anushakti Nagar, Mumbai 400 094, India*

[a)]Corresponding author: shekharpandey@rrcat.gov.in

**Abstract.** A study on the optimization of ambient temperature growth and superconducting properties of Ti-V alloy thin films grown on $SiO_2$-coated Si substrate is reported here. These films have been synthesized by co-sputtering of Ti and V targets, and films having different Ti concentrations were deposited to get the optimized critical temperature ($T_C$) of thin films close to the bulk value. The maximum $T_C$ of 5.2 K has been obtained in the $Ti_{40}V_{60}$ composition, which is further increased to 6.2 K when a 10 nm thick Ti underlayer is added below the Ti-V film. GIXRD measurements confirm the formation of Ti-V alloys in the desired crystal structure. The upper critical field ($H_{C2}$) of the thin films has been estimated with the help of magnetotransport measurements. The utility of Ti-V alloy thin films in superconducting radiation detection applications is ascertained.

## INTRODUCTION

Ti-V alloys are promising as an alternative to the bulk Nb and Nb-based superconductors for high-field applications, particularly in the neutron radiation environment [1, 2]. The Ti-V-based alloys are found to have a high critical current density ($J_C$) in zero field and in the presence of magnetic fields [3]. The superconducting transition temperature ($T_C$) of the Ti-V alloys is in the same temperature range as the Nb-based superconductors like the Nb-Ti alloys [4]. In addition to the bulk, thin films of these superconductors may be useful in various technological applications like radio frequency cavities, quantum electronics, and superconducting radiation detectors. Presently, there are very few reports on synthesising Ti-V alloy superconducting thin films [5, 6]. It is reported that the synthesis of Ti-V alloy thin films was either carried out at high temperatures [5] or at cryogenic temperatures [6]. The present thin films are synthesized for studying their potential as superconducting radiation detector elements. Hence, they are grown at ambient temperatures as this greatly simplifies the design and fabrication of the detectors.

Here, we report a study on the Ti-V alloy superconducting thin films synthesized at room temperature using DC magnetron co-sputtering technique. For the optimization of the films for relevant superconducting and normal state properties for possible radiation detector applications, they were deposited with different compositions and physical properties. The $T_C$ of the bulk Ti-V alloy system varies with the Ti and V content [7]. In the present work, to optimize the $T_C$ of the Ti-V alloy thin films, the Ti content is varied from 30% to 50%, as the bulk $T_C$ is reported to be higher in this composition range [7]. The aim was to achieve a $T_C$ close to the highest reported bulk value [3]. The films were characterized for phase purity, morphology and DC electrical transport properties. Magnetotransport measurements were also performed to estimate the upper critical field.

## EXPERIMENTAL DETAILS

### Synthesis of Ti-V alloy thin films

The alloy thin films of Ti and V were deposited using the co-sputtering technique, in a home-built DC magnetron sputtering system. The high-purity Ti (99.99 %) and V (99.9 %) targets were co-sputtered onto the $SiO_2$ (300 nm thick, amorphous)-coated Si (100) substrate (0.5 mm thick) in an ultra-high vacuum environment. To deposit the films with particular Ti : V ratios, the pre-calibrated individual deposition rates of Ti and V targets were used to calculate the molar volumes required for the final alloy composition. The deposition rates as a function of deposition time and current for the V and Ti targets were measured using a surface profilometer. To improve the grain size and $T_C$ of the films, some compositions were grown on a 10 nm thick Ti underlayer. Two of the four magnetron guns arranged in the confocal geometry were DC powered simultaneously for the growth of alloy thin films. A load lock chamber is attached to the main deposition chamber to avoid contamination during the substrate transfer. $10 \times 10$ mm$^2$ substrates were ultrasonically cleaned in de-ionized water, boiling acetone and ethyl alcohol and then dried at 100$^0$C before loading in the substrate holder. To ensure the homogeneity of the films, the substrates were rotated at 15 RPM in the focal plane of the magnetron guns. The thin films were deposited at room temperature, under the background argon gas pressure of $2\times10^{-3}$ mbar by controlling the gas inflow using a mass flow controller. Prior to the deposition, the main deposition chamber was pumped down to less than $1 \times 10^{-7}$ mbar pressure. To deposit the films of different compositions, sputtering of the V and Ti targets was performed with varying sputtering currents.

## Structural and electrical characterization

Grazing incidence X-ray diffraction (GIXRD) measurements were performed for the confirmation of phase purity and the thickness of the films was measured using X-ray reflectivity measurements using an X-ray diffractometer (Bruker D8 Discover) employing CuKα radiation. The high-resolution images of the films were taken using a Sigma Carl-Zeiss Scanning Electron Microscope (SEM) and the composition of the films was confirmed by energy dispersive x-ray analysis (EDX).

Measurement of electrical resistance (R) as a function of temperature (T) was performed in the T range 2-300 K using a Cryo-free Spectromag CFSM7T-1.5 magneto-optical cryostat system (Oxford Instruments, UK), with the probes configured in the Van der Pauw four-probe geometry. High-conductivity silver paint was used to make the electrical connections of thin copper wires with the samples.

For the estimation of the upper critical field of the films ($H_{C2}$), their electrical resistance was measured as a function of T under constant applied magnetic fields up to 4.5 T. We determined the $H_{C2}$ value at 90% of the normal state resistance on the temperature-dependent resistance curve and then plotted the $\mu_0H_{C2}$ values as a function of T. For the estimation of $H_{C2}$ at 0 K we have fitted the transition temperature $T_C(\mu_0H)/T_C(0)$ versus the applied field $\mu_0H$ curve using functional form

$$\mu_0 H_{C2}(T) = \mu_0 H_{C2}(0) \left[1 - \left(\frac{T}{T_C}\right)^2\right] \qquad (1)$$

## RESULTS AND DISCUSSION

V and Ti contents in the Ti-V alloy thin films were controlled by varying the sputtering currents for the V and Ti targets, and Ti$_x$V$_{100-x}$ alloy films were deposited for x = 30, 36, 40, 43 and 50. Since the sputtering currents were varied to get the desired film compositions, the deposition times were adjusted to keep the film thickness constant at 45 nm. The phase purity of the films was studied using GIXRD plots of normalized intensity versus 2θ shown in fig1(a). All the peaks have been indexed to the bcc crystal structure (space group Im$\bar{3}$m). Peaks corresponding to individual Ti and V phases were not observed, which indicates the compositional homogeneity of the films. It is observed in fig. 1(a) that the peaks shift towards the lower 2θ values as the Ti content in the films increases. The average grain sizes of the V and Ti$_x$V$_{100-x}$ films were estimated using the Williamson-Hall plot to be 10-15nm.

Fig. 1(b) shows the high-resolution SEM image of the Ti$_{40}$V$_{60}$ alloy thin film. The image shows some white patches on the surface of the film. The surface morphology of the present films is similar to that reported for the cathodic arc-deposited Ti-V films [8]. Signatures of grain boundaries could not be seen in the images indicating that the grain sizes are smaller than the SEM resolution. The elemental and compositional quantification of Ti-V alloy films was done by the EDAX measurements and the estimated compositions were found to be close to the nominal values.

The R(T) curves for the Ti-V alloy thin films in the T range 2- 300 K shown in fig. 2 confirm the superconducting nature of the films. The highest $T_C$ ~ 5.2 K is observed in the Ti$_{40}$V$_{60}$ composition. The normalized resistance as a function of temperature is shown in fig.2(a). To further improve the $T_C$ of the films, the Ti-V alloy thin films were also deposited with a 10 nm thick Ti underlayer. Two different compositions Ti$_{40}$V$_{60}$ and Ti$_{30}$V$_{70}$ representing the highest achieved $T_C$ and lowest Ti content respectively were chosen for the growth on the Ti underlayer. A significant improvement in the $T_C$ (> 6 K) has been observed in the Ti$_{40}$V$_{60}$ film grown on the Ti underlayer. This increase in the $T_C$ of the films with Ti underlayer seems to be related to the increase in the average grain sizes of the Ti-V films which can be attributed to the reduction in the oxygen partial pressure during deposition, as Ti is a getter material that absorbs oxygen [9]. Less oxide formation improves the grain growth in the films, which leads to the improvement of $T_C$.

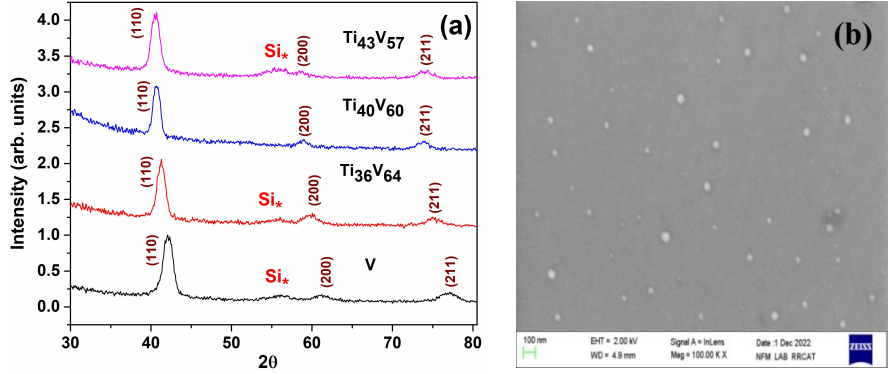

**FIGURE 1. (a)** GIXRD patterns of different compositions of Ti-V alloy thin films. A systematic shift of the peaks towards lower $2\theta$ values is seen for $Ti_xV_{100-x}$ (x = 0, 36, 40, 43) compositions. **(b)** High-resolution SEM image of the $Ti_{40}V_{60}$ alloy thin film.

Fig.2(b) is the normalized resistance as a function of temperature for Ti-V alloy thin films with 10 nm Ti underlayer compared with the normalized R(T) of the film having highest $T_C$ but without Ti underlayer. It can be seen from fig.2(a) and fig.2(b) that all the Ti-V alloy thin films exhibit a superconducting transition to a zero-resistance state and the corresponding $T_C$ values are presented in table 1.

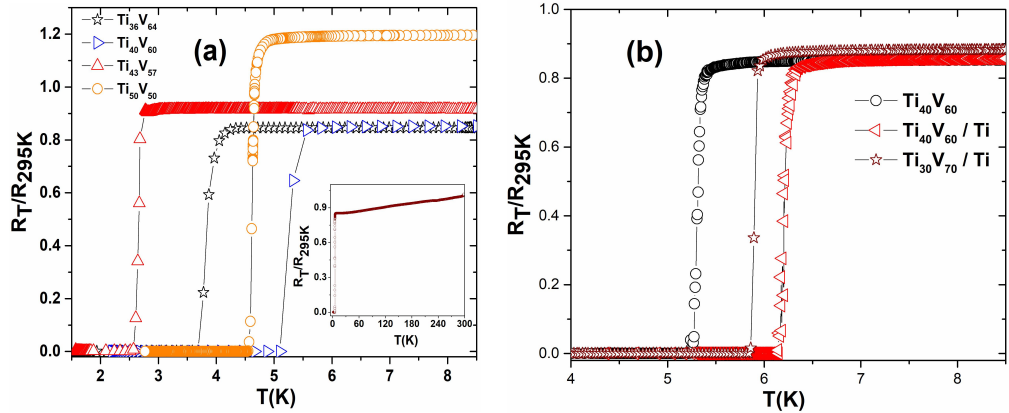

**FIGURE 2. (a)** Resistance as a function of temperature for $Ti_xV_{100-x}$ (x = 36,40,43,50) thin films. Inset: R(T) of $Ti_{40}V_{60}$ film in full temperature range 2- 300 K. **(b)** Resistance as a function of temperature for $Ti_xV_{100-x}$ (x = 30,40) alloy thin films with 10 nm Ti under layer ($Ti_xV_{100-x}$ / Ti) and film of similar composition without Ti underlayer.

**TABLE 1. Superconducting transition temperatures of TiV alloy thin films**

| S.No. | Composition | $T_C$ (K) |
|---|---|---|
| 1. | $Ti_{36}V_{64}$ | 3.8 |
| 2. | $Ti_{40}V_{60}$ | 5.2 |
| 3. | $Ti_{43}V_{57}$ | 2.6 |
| 4. | $Ti_{50}V_{50}$ | 4.6 |
| 5. | $Ti_{40}V_{60}$ (with 10 nm Ti underlayer) | 6.2 |
| 6. | $Ti_{30}V_{70}$ (with 10 nm Ti underlayer) | 5.9 |

The results of magnetotransport measurements performed in fields up to 4.5 T on the $Ti_{36}V_{64}$ sample are presented in fig. 3. With increasing magnetic fields, the $T_C$ of the film decreases. Estimation of the upper critical field at 0 K ($H_{C2}(0)$) was done by fitting the $T_C$ versus $\mu_0 H$ curve using equation (1), keeping $H_{C2}(0)$ as a fitting parameter [fig.3(b)]. The fitted value of the $\mu_0 H_{C2}$ at T = 0 is found to be 5.3 (± 0.1) T.

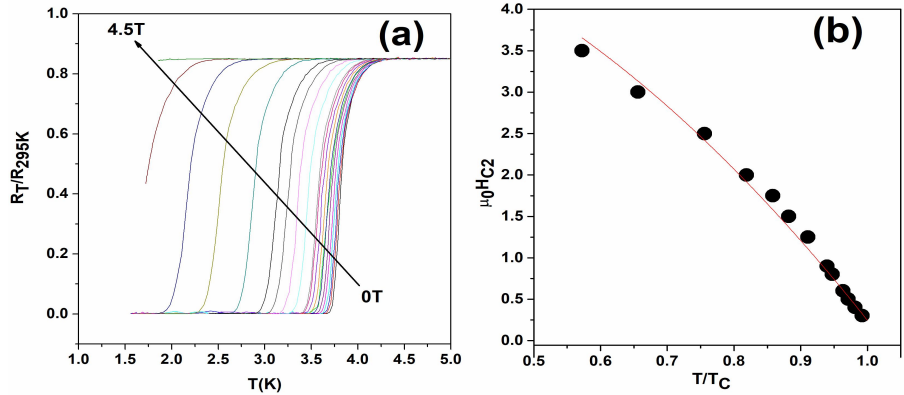

**FIGURE 3. (a)** Resistance as a function of temperature, R(T), for $Ti_{36}V_{64}$ alloy thin film, measured in different applied magnetic fields. **(b)** Critical field $H_{C2}$ as a function of reduced temperature $T/T_C$; the red curve indicates the fitting of the $H_{C2}$ using equation (1).

## CONCLUSIONS

In conclusion, we have successfully synthesized and characterized the Ti-V alloy thin films with $Ti_xV_{100-x}$ (x = 30, 36, 40, 43, 50) compositions and achieved a $T_C$ value of 6.2 K in the $Ti_{40}V_{60}$ composition with 10 nm Ti under layer which is quite close to its bulk value. Structural, electrical and magneto-transport studies have been performed on the films and the results confirm the phase purity and elemental composition of the films. The $H_{C2}(0)$ of the $Ti_{36}V_{64}$ alloy estimated from magnetotransport measurement is 5.3 (± 0.1) T. The good quality Ti-V alloy films with high $T_C$ and $H_{C2}(0)$ are good candidate materials for the fabrication of cryogenic radiation detectors for the Mid and Far IR range.

## ACKNOWLEDGEMENT

The authors are thankful to Dr. Sanjay Rai and Dr. Pooja Gupta for the GIXRD experiment and Smt. Rashmi Singh for the SEM-EDX measurements. Shekhar thanks HBNI-RRCAT for the financial support.

## REFERENCES

1. M. Tai, K. Inoue, A. Kikuchi, T. Takeuchi, T. Kiyoshi, and Y. Hishinuma, *IEEE Trans. Appl. Supercond*. 17, 2542 (2007).
2. S. T. Sekula, *J. Nucl. Mater*. 72, 91 (1978).
3. Sabyasachi Paul, SK. Ramjan, R. Venkatesh, L. S. Sharath Chandra, and M. K. Chattopadhyay, *IEEE Trans. Appl. Supercond.,* 31, 8000104 (2021).
4. Y. Shimizu, K. Tonooka, Y. Yoshida, M. Furuse1, and H. Takashima, *Sci Rep* 8, 15135 (2018)
5. H. J. Spitzer, *Low Temperature Physics-LT 13,* 485 (1974).
6. B. Bandyopadhyay, P. Watson, Y. Bo, and D. G. Naugle, *J. Phys. F: Met. Phys*. 17, 433 (1987).
7. E.W. Collings, P.E. Upton, and J.C. Ho, *J. Less-Common Met.* 42 , 285 (1975).
8. Yin-Yu Chang, Jia-Hao Zhang, and Heng-Li Huang, *Materials.* 11, 2495 (2018).
9. Werner Espe, Max Knoll, and Marshall P. Wilder, *Getter Materials. Electronics* 23, 80 (1950).